# Baryon Acoustic Oscillations in the Lyman Alpha Forest

**Michael L. Norman, Pascal Paschos & Robert Harkness**

Laboratory for Computational Astrophysics, SERF 0424, University of California San Diego, 9500 Gilman Drive, La Jolla, CA 92093

mlnorman@ucsd.edu

**Abstract**. We use hydrodynamic cosmological simulations in a (600 Mpc)$^3$ volume to study the observability of baryon acoustic oscillations (BAO) in the intergalactic medium as probed by Lyman alpha forest (LAF) absorption. The large scale separation between the wavelength of the BAO mode (~150 Mpc) and the size of LAF absorbers (~100 kpc) makes this a numerically challenging problem. We report on several $2048^3$ simulations of the LAF using the ENZO code. We adopt WMAP5 concordance cosmological parameters and power spectrum including BAO perturbations. 5000 synthetic HI absorption line spectra are generated randomly piercing the box face. We calculate the cross-correlation function between widely separated pairs. We detect the BAO signal at z=3 where theory predicts to moderate statistical significance.

## 1. Introduction

Baryon Acoustic Oscillations (BAO) is a promising standard ruler technique in observational cosmology to measure the expansion history of the universe and thereby a probe of cosmic dark energy [1]. BAO refers to acoustic waves in the cosmic photon-baryon plasma prior to decoupling at z=1080. They have a characteristic wavelength which is the sound speed in the plasma ~$c/3^{1/2}$ times the age of the universe at decoupling. For the concordance cosmological model this is about 150 comoving Mpc [2].

BAO modulates the power spectrum of density fluctuations that seed structure formation. Modes of this wavelength have been detected in the cosmic microwave background (CMB) anisotropies and galaxy large scale structure (LSS)[1]. In principle, the BAO should be imprinted on all forms of cosmic structure including the distribution of intergalactic gas as probed by the Lyman alpha forest (LAF).

The LAF is a thicket of narrow absorption lines seen in the spectra of high redshift quasars [3]. It is caused by resonant Ly $\alpha$ absorption in intergalactic hydrogen clouds spread out in wavelength by the cosmological redshift. A high resolution quasar absorption line spectrum thus probes the detailed spatial distribution of intergalactic gas along a sightline many hundreds of Mpc in length. The best way to detect BAO in the LAF is by cross-correlating absorption spectra in widely separated quasar pairs. To date, the BAO has not been detected in the LAF for lack of sufficient data (it is a statistical detection; see below), although surveys are being designed to remedy this [4]. [5] have used analytic arguments to estimate how much data might be needed to reach the precision required in dark energy studies. In this work we use hydrodynamical cosmological simulations of unprecedented size to explore the eventual detectability of BAO in the LAF.

## 2. Simulations

We have used the hydrodynamic cosmological code ENZO [6] to simulate the structure of the IGM on scales exceeding the BAO length scale. The large scale separation between the wavelength of the fundamental BAO mode (~150 Mpc) and the size of LAF absorbers (~100 kpc) makes this a numerically challenging problem. We have done two simulations at $2048^3$ resolution in cubic computational domains 307 and 614 Mpc on a side. The cosmological parameters were taken to be the 5 –year WMAP concordance parameters: $\Omega_b$=0.043, $\Omega_{dm}$=0.207, $\Omega_\Lambda$=0.75, h=0.72, $\sigma_8$=0.8, n=0.95. The first three are the fraction of the closure density in baryons, dark matter, and vacuum energy, respectively; h is the Hubble parameter in units of 100 km/s/Mpc; and $\sigma_8$ and n are the amplitude and primordial slope of the matter fluctuation power spectrum, respectively. Gaussian random field initial conditions were generated at the starting redhsift z=60 in the usual way using an input power spectrum obtained from CAMB08 [7] including the BAO features.

ENZO evolves both the dark matter and primordial hydrogen and helium gas in a cubic, triply periodic volume subject to self-gravity in an expanding universe. A photoionizing UV background radiation field is assumed which controls the ionization state of the gas. The equations of multispecies gas dynamics are evolved on a uniform Cartesian grid of size $2048^3$ cells using the PPM method. This yields a co-moving grid resolution of 150(300) kpc for the smaller(larger) box. Dark matter is represented as $2048^3$ collisionless particles, and is evolved using the particle-mesh technique. The Poisson for the self-consistent gravitational field is solved using FFTs. ENZO is parallelized for execution on distributed memory clusters with multicore nodes using a combination of MPI and OpenMP. See [6] for more details on the ENZO code.

## 3. Results

A volumetric rendering of the logarithm of the gas density at z=3 from the 300 Mpc simulation is shown in Fig. 1a. One recognizes the characteristic filament/void "cosmic web" structure imprinted by the clustering of cold dark matter [8]..Lines of sight intersecting the baryonic filaments create the LAF absorption seen in the spectra of distant quasars. Modulation of the cosmic web by BAO is too subtle to be detected by eye, so statistical techniques are employed.

To this end synthetic HI absorption spectra are generated using the method described in [9] for N=5000 lines of sight randomly piercing a face of the cube and running parallel to a coordinate axis. A sample spectrum is shown in Fig. 1b. Here the black line is the spectrum at the native resolution of the simulation (corresponding to a spectral resolution of 13000), and the red line is the spectrum convolved with the resolution of the SDSS spectrometer (spectral resolution 1800). The cross-correlation function among the N(N-1)/2 pairs is calculated and binned versus comoving separation on the sky (equivalent to angular separation). The result is shown in Fig. 2a. Here we have used the highest resolution spectra and ignore noise. The $1\sigma$ standard error to the mean of the cross correlation in 1 Mpc bins including the propagation of error due to the variance in the product of two sightlines is shown as error bars. The BAO signal is clearly evident at the predicted separation of 155 comoving Mpc. We estimate the statistical significance of this result to be $1.86\sigma$ by comparing the amplitude of the peak, after we subtract the upper level of error at the zero correlation level, to the size of the error bar at the peak. The bump at comoving separation of ~75 Mpc appears to be unrelated to the BAO phenomenon and is statistically insignificant within our margin of error. Note that the statistical significance depends on our sample size of Ly $\alpha$ skewers through the volume. The variance of the sightline pair product will not change using a larger sample, but the error to the mean per correlation bin (x) will decrease as $Np(x)^{-0.5}$, where Np(x) is the number of pairs per comoving separation.

In Fig 2b we explore the physical origin of the BAO signal. We plot the cross-correlation function of dark matter density, baryon density, and baryon temperature along same lines of sight used for Fig. 2a. We see a narrow peak in all three quantities at the BAO distance scale (155 Mpc). Temperature shows the strongest signal, followed by the dark matter and baryon densities at comparable amplitudes. The shoulder to the right of the BAO peak we interpret as due to a local maximum in the fluctuating signal at 170 Mpc; other peaks can be seen at separations of 75, 125, and 200 Mpc.

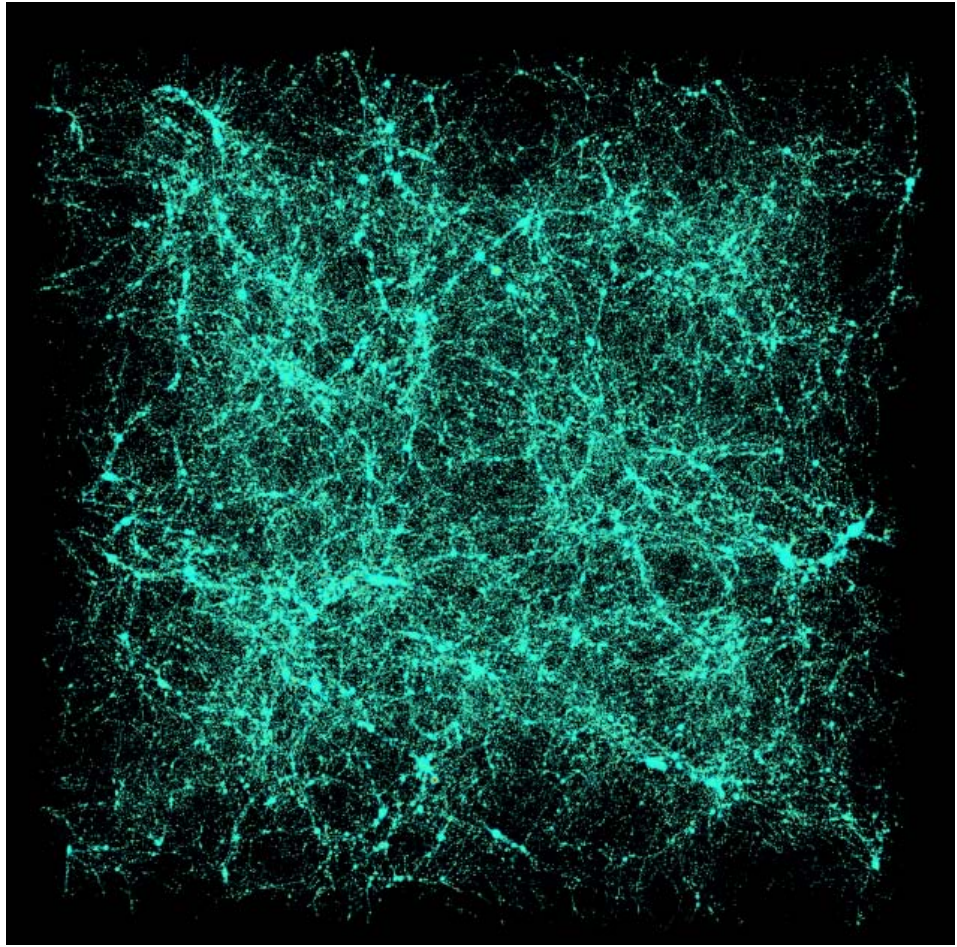

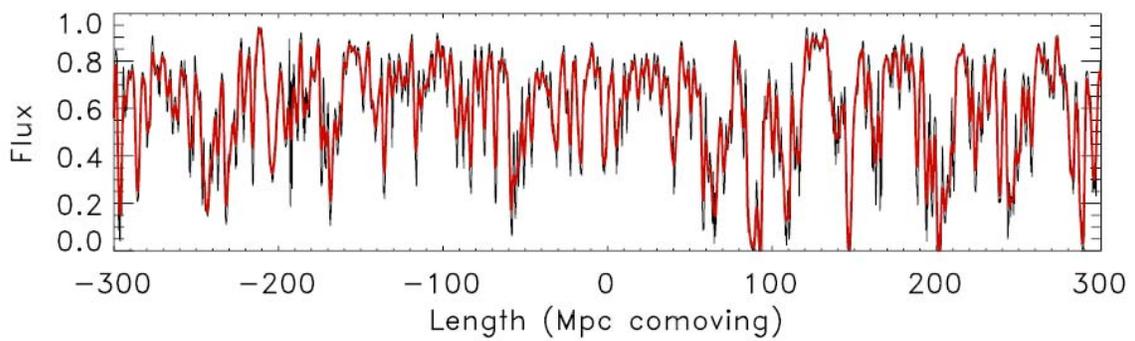

**Fig. 1:** (a) $2048^3$ hydrodynamic simulation of the Lyman alpha forest in a 307 Mpc cube of the universe. Rendering of log(baryon density), courtesy J. Insley, ANL. (b) synthetic neutral hydrogen absorption spectrum for one LOS through a 614 Mpc cube. Black and red lines are at spectral resolutions of 13,000 and 1800, respectively.

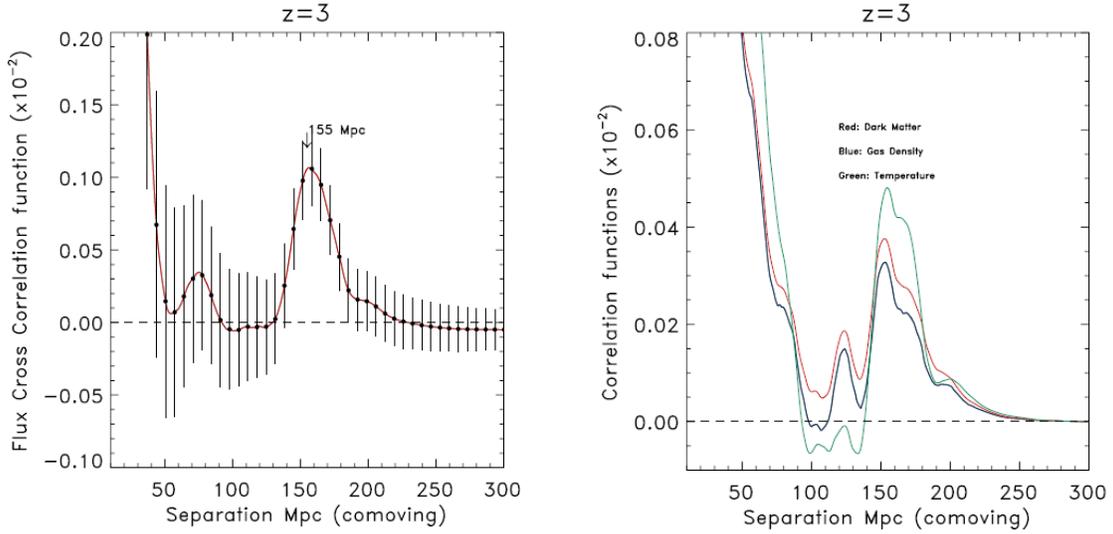

**Fig. 2:** (a) cross correlation of transmitted flux at z=3 in quasar pairs versus separation on the sky, in comoving Mpc. (b) cross correlation of dark matter density, baryon density, and baryon temperature along same lines of sight used for Fig. 2a. The BAO signal is evident at ~150 Mpc.

## 4. Discussion and Future Work

We have simulated the observability of baryon acoustic oscillations (BAO) in the Lyman alpha forest using hydrodynamic cosmological simulations in large boxes (300 and 600 Mpc on a side) and high resolution ($2048^3$ particles and cells). With a wavelength of ~150 Mpc (comoving), the BAO is barely(easily) detectable in the 300(600) Mpc volumes. At a redshift of z=3, the BAO signal stands out in the cross correlation function of the transmitted flux in widely separated simulated quasar absorption line spectra. The signal is also present in skewers of dark matter density, and baryon density and temperature.

Restricting our attention to results from the 600 Mpc volume (Fig. 2), the statistical significance of the detection is about $1.86\sigma$. However we have used ideal spectra in obtaining this result. In the future it will be important to study the effects of spectral resolution and noise on the BAO signal. We are currently repeating the calculation of the flux cross correlation function lowering the spectral resolution to that of the upcoming BOSS survey [4].

It is also important to study the effect of redshift evolution. The BAO signal should increase with decreasing redshift as the BAO mode grows in amplitude and the LAF absorption becomes less saturated. However, below z=2 the intergalactic LAF absorption begins to disappear while the circum-galactic absorption becomes more prominent [10]. The redshift at which the BAO signal is maximal is not known at the present time, but should be easily estimated with numerical simulations.

Finally, we also need to quantify the effect of numerical resolution on the BAO signal. At $2048^3$ cells, our spatial resolution is only 300 kpc in the 600 Mpc volume. This is relatively poor resolution when it comes to resolving the internal structure of baryonic features which give rise to the LAF [11]. We are currently running a $4096^3$ simulation in a 600 Mpc volume which should significantly improve the measurement and our understanding of the numerical systematics.

**Acknowledgements**
The authors acknowledge helpful discussions with Martin White. We wish to thank Joe Insley at Argonne National Lab for assistance with the rendering in Fig. 1. MLN and PP acknowledge partial support by NSF grants AST-0708960 and AST-0808184. Simulations were carried out on NERSC Seaborg under the INCITE program and NICS Kraken under LRAC allocation TG-MCA98N020.

**References**

[1] Eisenstein, D. and Bennett, C. Physics Today, April 2008, pp. 44-50; Eisenstein, D. et al. 2007, Astrophysical J., Vol. 633, p. 560
[2] Bashinsky,S.; Bertschinger, E. 2001. Physical Review Letters, vol. 87, Issue 8
[3] Rauch, M. 1998. Annual Review of Astronomy and Astrophysics, Volume 36, pp. 267-316.
[4] Schlegel, D. J.; Blanton, M.; Eisenstein, D.; Gillespie, B.; Gunn, J.; Harding, P.; McDonald, P.; Nichol, R.; Padmanabhan, N.; Percival, W.; and 8 coauthors, 2007. Bulletin of the American Astronomical Society, Vol. 39, p. 966
[5] McDonald, P. & Eisenstein, D. 2007. Physical Review D, Vol. 76, Issue 6
[6] Norman, M.L.; Bryan, G.L.; Harkness, R.; Bordner, J.; Reynolds, D.; O'Shea, B.; Wagner, R. 2007. in Petascale Computing: Algorithms and Applications, Ed. D. Bader, CRC Press LLC
[7] A. Lewis, A. Challinor, and A. Lasenby 2000. The Astrophysical Journal Vol. 538, p. 473, http://camb.info, astroph-ph/9911177
[8] Bond, J. R., Kofman, L., Pogosyan, D. 1996. Nature, Vol. 380, pp. 603
[9] Zhang, Y., Anninos, P., Norman, M. L., and Meiksin, A. 1997. The Astrophysical Journal Vol. 485, p.496
[10] Davé, R.; Hernquist, L.; Katz, N.; Weinberg, D. H. 1999. The Astrophysical Journal, Volume 511, Issue 2, pp. 521-545.
[11] Jena, T., Norman, M.L., Tytler, D.T., Kirkman, D., Suzuki, N., Chapman, A., Melis, C., Paschos, P., O'Shea, B, So, G. and 5 coauthors, 2005. Monthly Notices of the Royal Astronomical Society, Volume 361, Issue 1, pp. 70-96.